\documentclass[twocolumn]{aastex631}

\begin{document}

\renewcommand{\topfraction}{1.0}
\renewcommand{\bottomfraction}{1.0}
\renewcommand{\textfraction}{0.0}

\newcommand{\kms}{km~s$^{-1}$\,}
\newcommand{\masyr}{mas~s$^{-1}$\,}
\newcommand{\msun}{$M_\odot$\,}

\title{Exploring Thousands  of Nearby  Hierarchical Systems  with Gaia
  and Speckle Interferometry}

\shorttitle{Thousands of Nearby Hierarchies with Gaia and Speckle}

\author{Andrei Tokovinin}
\affiliation{Cerro Tololo Inter-American Observatory --- NSF's NOIRLab
Casilla 603, La Serena, Chile}
\email{andrei.tokovinin@noirlab.edu}

\begin{abstract}
There  should be  about  10,000 stellar  hierarchical systems  within
100\,pc with primary stars more massive  than 0.5 \msun, and a similar
amount  of  less  massive  hierarchies. A  list  of  8000  candidate
multiples is derived  from wide binaries found in the  Gaia Catalog of
Nearby Stars where  one or both components  have excessive astrometric
noise  or other  indicators of  inner  subsystems.  A  subset of  1243
southern candidates were observed with  high angular resolution at the
4.1 m telescope, and 503 new  pairs with separations from 0\farcs03 to
1\arcsec ~were resolved. These data allow estimation of the inner mass
ratios and periods and help to  quantify the ability of Gaia to detect
close pairs.   Another 621 hierarchies  with known inner  periods come
from the Gaia  catalog of astrometric and  spectroscopic orbits. These
two non-overlapping  groups, combined  with  existing ground-based
data,  bring the  total number  of known  nearby hierarchies  to 2754,
reaching  a completeness  of  $\sim$22\% for  stars  above 0.5  \msun.
Distributions of their periods and  mass ratios are briefly discussed,
and the prospects of further observations are outlined.
\end{abstract}

   \keywords{binaries:visual}


\section{Introduction}
\label{sec:intro}

Stars form in groups. Almost every star has been gravitationally bound
to some  other star or stars  in their infancy \citep{Lee2019},  and a
substantial fraction of  these systems have survived,  as evidenced by
the    multiplicity   statistics    of   mature    field   populations
\citep{Moe2017,Offner2022}.  Statistics  of stellar  systems helps
us to  understand their  formation and early  evolution.  Hierarchical
systems are  particularly informative in this  regard.  However, owing
to  the  vast range  of  parameters  (separations, mass  ratios),  the
complete view  of even the  nearest population of  stellar hierarchies
has been  difficult to  grasp. The relatively  well studied  sample of
solar-type stars  within 25\,pc contains only  56 hierarchical systems
\citep{R10}.  

The   Gaia   astrometric   space   mission   \citep{Gaia1,Gaia3}   has
dramatically changed the landscape of  Galactic astronomy in many ways.
The mission  continues, and the use  of its intermediate data  for the
study   of    stellar   systems    is   a   rapidly    growing   field
\citep[e.g.][]{ElBadry2021,Tokovinin2022b}.  The Gaia  Catalog of Nearby Stars
(GCNS)  within 100\,pc  \citep{GCNS},  based on  the  Gaia Early  Data
Release 3  (eDR3), gives a  complete census of  all stars down  to the
hydrogen  burning limit  (except for some  binaries lacking  parallaxes).
Owing  to  its exquisite  astrometric  precision,  Gaia can  detect  a
substantial fraction of binary systems in the 100 pc volume.  However,
the periods  and  mass ratios  of  most  candidate close  binaries  remain
essentially   unconstrained.   The   Non-Single  Star   (NSS)  catalog
\citep{Arenou2022,Pourbaix2022},  part  of  the Gaia  data  release  3
(DR3),  contains  orbital  elements  only  for  a  small  fraction  of
astrometric and spectroscopic binaries detected by Gaia.
 
In this  work, I open  the treasure  trove of   Gaia data to  get a
better view of  nearby stellar hierarchies.  A candidate list
is created  by isolating bound  pairs of stars  found in the  GCNS and
looking at those  that contain signs of inner  subsystems according to
the Gaia  binarity indicators.   Naturally, some of  these hierarchies
are already known from prior work. A subset of the new candidates have
been observed  systematically by speckle  interferometry at the  4.1 m
Southern  Astrophysical Research  Telescope (SOAR)  in 2021--2023,  and
these results  are reported here.   About half of the  candidates were
resolved, providing estimates of their mass ratios and likely periods.
At the same time, these  resolutions allow better understanding of the
discovery potential of the Gaia binarity indicators. Complementing the
known hierarchies by  these new systems and by the  systems with inner
orbits  determined by  Gaia leads  to a  sample of 2758 main-sequence
hierarchies within  100\,pc with  known or  estimated inner  and outer
periods.   Their  primary  stars   are  generally  more  massive  than
$\sim$0.7 \msun.  A glimpse of  their statistics (still incomplete but
much better  than in  the pre-Gaia  era) is  given, and  directions of
future research are outlined.

\section{Gaia Hierarchies within 100 pc}
\label{sec:sample}

\subsection{Number of Hierarchies within 100 pc}

The GCNS  \citep{GCNS} is a rich  source of hierarchical systems  in a
volume-limited sample, with  the potential to make  a major contribution
to their statistics.  It contains 331,312 entries. However, the GCNS misses
close binaries  with components of  comparable brightness that  do not
have parallaxes in eDR3.  The  fraction of missing stars was estimated
at 7.4\%  in the GCNS,  based on its data  for  10 pc  volume.  The
peak of binary  separation distribution at $\sim$50  au corresponds to
an angular  separation of  0\farcs5 at 100\,pc,  so the  bias against
binaries  in the  complete  GCNS could  be larger  than  in its  10 pc
portion.  Empirical characterization of the Gaia bias against binaries
based on the new SOAR observations is provided below.

\begin{figure}
\epsscale{1.1}
\plotone{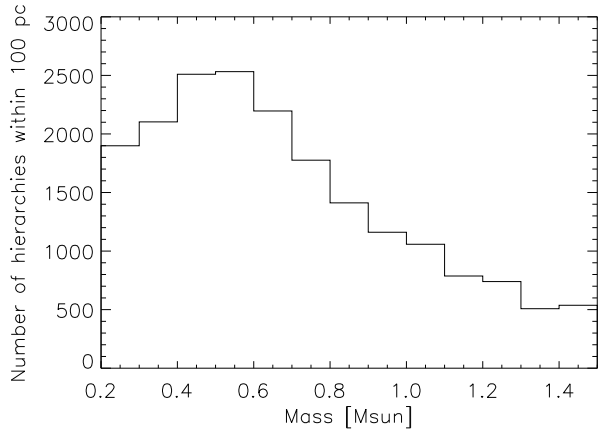}
\caption{Estimated number of hierarchical systems within 100\,pc in
  0.1 \msun mass bins.
\label{fig:trifrac}
}
\end{figure}

The distribution  of absolute magnitudes in  the GCNS peaks at  $M_G =
10.5$ mag,  reaching a density of  0.01 star~pc$^{-3}$~mag$^{-1}$, and
drops smoothly on  both sides of the maximum (see  Figure~16 in GCNS).
The corresponding  median mass is  0.32 \msun according to  the PARSEC
isochrone for  solar metallicity  \citep{PARSEC}. Let us  estimate how
many hierarchical  (e.g.  triple  and higher-order) systems  are there
within 100\,pc.   It is  well known that  the fraction  of hierarchies
increases  with  mass,  while  the   density  of  stars  declines.   I
approximated  the fraction  of hierarchies  vs.  mass  in Figure~1  of
\citet{Offner2022} by a parabola $f_H \approx  0.146 + 0.255 x + 0.414
x^2$, where $x = \log_{10} M/$\msun,  and multiplied the star counts in
GCNS  by this  fraction (the  masses are  estimated from  the absolute
magnitudes $M_G$).  The result  in Figure~\ref{fig:trifrac} suggests a
total number around 19,000 for  masses above 0.2 \msun (10,400 above
0.5 \msun and 6,400 above  0.7 \msun).  Additional 4,400 hierarchies
are  predicted in  the first  0.1--0.2 \msun  bin, although  $f_H$ for
low-mass stars  is poorly known.   This model yields  4,062 hierarchies
with masses from 0.8 to 1.25 \msun within 100\,pc, roughly matching 56
systems found in the  64$\times$ smaller  volume by \citet{R10}.

\subsection{Selection of Candidate Hierarchies}

The GCNS provides a  list of 19,176 pairs of stars,  16,556 of which are
estimated to be bound. However, inner subsystems in triples bias  Gaia
measurements of parallaxes and proper motions (PMs), so many wide
pairs with subsystems  appear as unbound or even unrelated.  To avoid
potential bias against triples, I use the weaker criteria for selecting
 outer pairs, as outlined in \citet{Tokovinin2022b}: 
\begin{itemize}
\item
Parallaxes equal within 1\,mas.
\item
Projected separation $s < 20$ kau.
\item
Relative projected  speed (in  \kms) $\Delta  V <  10 (10^3/s)^{0.5}$,
where $s$ is expressed in au.  This is a relaxed form of the boundness
criterion  which   rejects  optical   companions  but   preserves  the
hierarchies. A similar approach was adopted by \citet{Hwang2020}.

\end{itemize}

A search over  GCNS with these criteria returns  24,604 systems, each
containing from  2 to 5 stars  (50,243 stars in total).  Most  systems are
just  wide  binaries,  except  944  triples,  42  quadruples,  and  one
quintuple, $\xi$~Sco.   The relaxed criteria  give a larger  sample of
wide systems compared to the list  of binaries given in the GCNS.  The
median mass of  stars in our wide  pairs is 0.44 \msun  (0.60 and 0.31
\msun for the primary and  secondary components, respectively).  This is
larger than the median mass in GCNS because binaries prefer stars more
massive than average  (in other words, the binary  fraction increases with
mass).   Other catalogs  of  wide  binaries based  on  Gaia have  been
published  by  \citet{Hartmann2020,ElBadry2021,Zavala2022}, and  others
using a variety of approaches and selection criteria.

Each  Gaia  (and GCNS)  entry  contains  two powerful  diagnostics  of
unresolved binaries, namely  the reduced unit weight  error (RUWE) and
the fraction  of double transits,  FDBL ({\tt IPDfmp} in  the Gaia
terminology).  I  also  explored  another  parameter,  {\tt  IPDgofha}
(an asymmetry parameter in  the Gaia image analysis), but found  it to be
poorly correlated with  RUWE and FDBL; for this  reason, probably, the
GCNS  does  not  contain  this  parameter. So,  FDBL,  RUWE,  and  the
variability  of radial  velocity (RVERR)  are the  main indicators  of
close  binaries.  Photometric  variability caused  by eclipses  is yet
another   indicator,    used   in   some   studies    of   hierarchies
\citep[e.g.][]{Hwang2020,Fezenko2022} but not relevant for this work.

It  is   generally  assumed  that  RUWE$>$1.4   indicates  significant
deviations from  a single-star  astrometric solution,  suggesting an
unresolved binary \citep{Belokurov2020,Penoyre2022}.  However, a large
RUWE can  be caused either by  the genuine motion of  the photocenter
(i.e.  an  astrometric binary) or by  the influence of a  faint visual
companion  that  spoils the  Gaia  astrometry  by its  presence;  both
situations occur and can be illustrated by concrete examples.

\begin{figure}
\epsscale{1.1}
\plotone{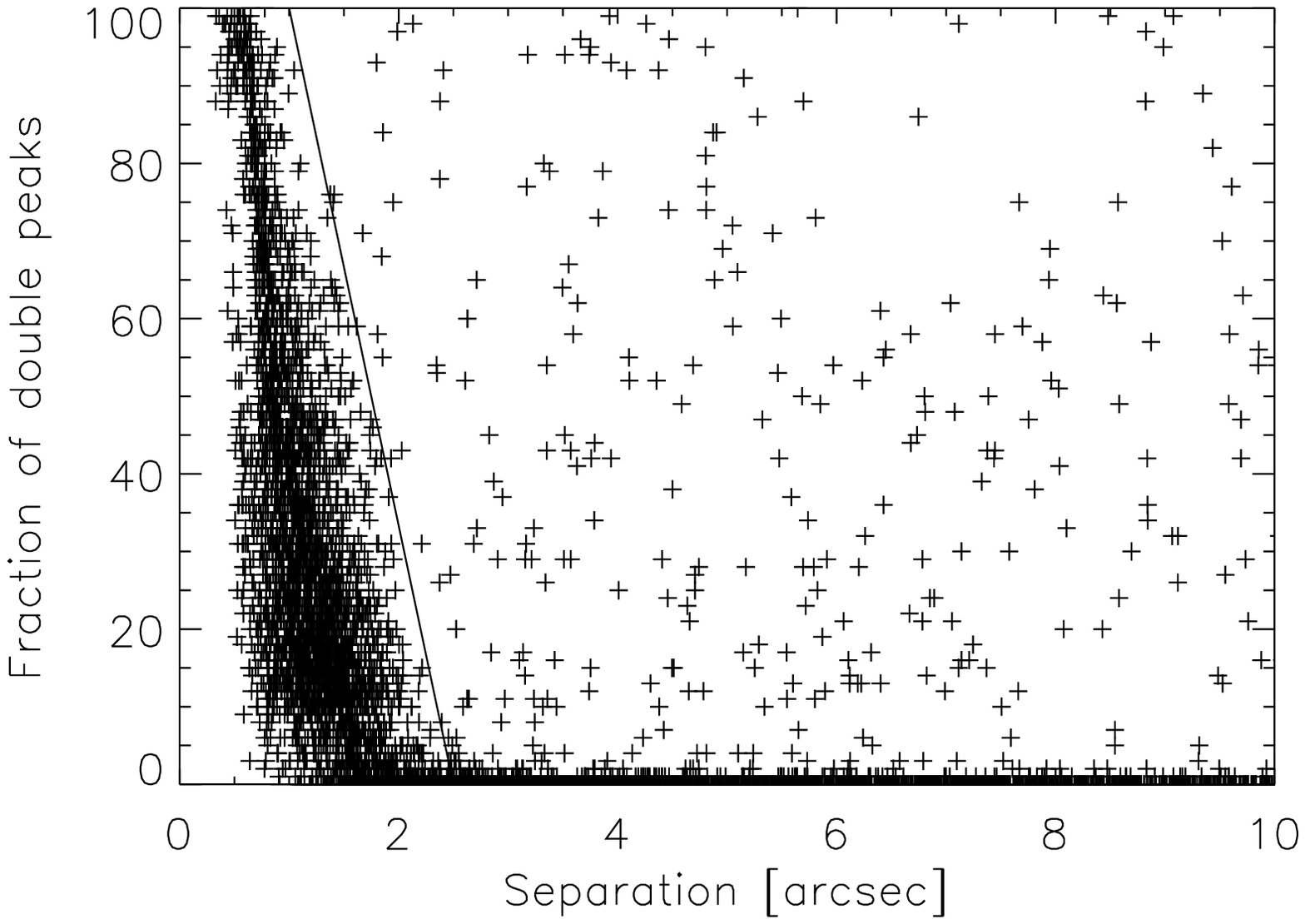}
\plotone{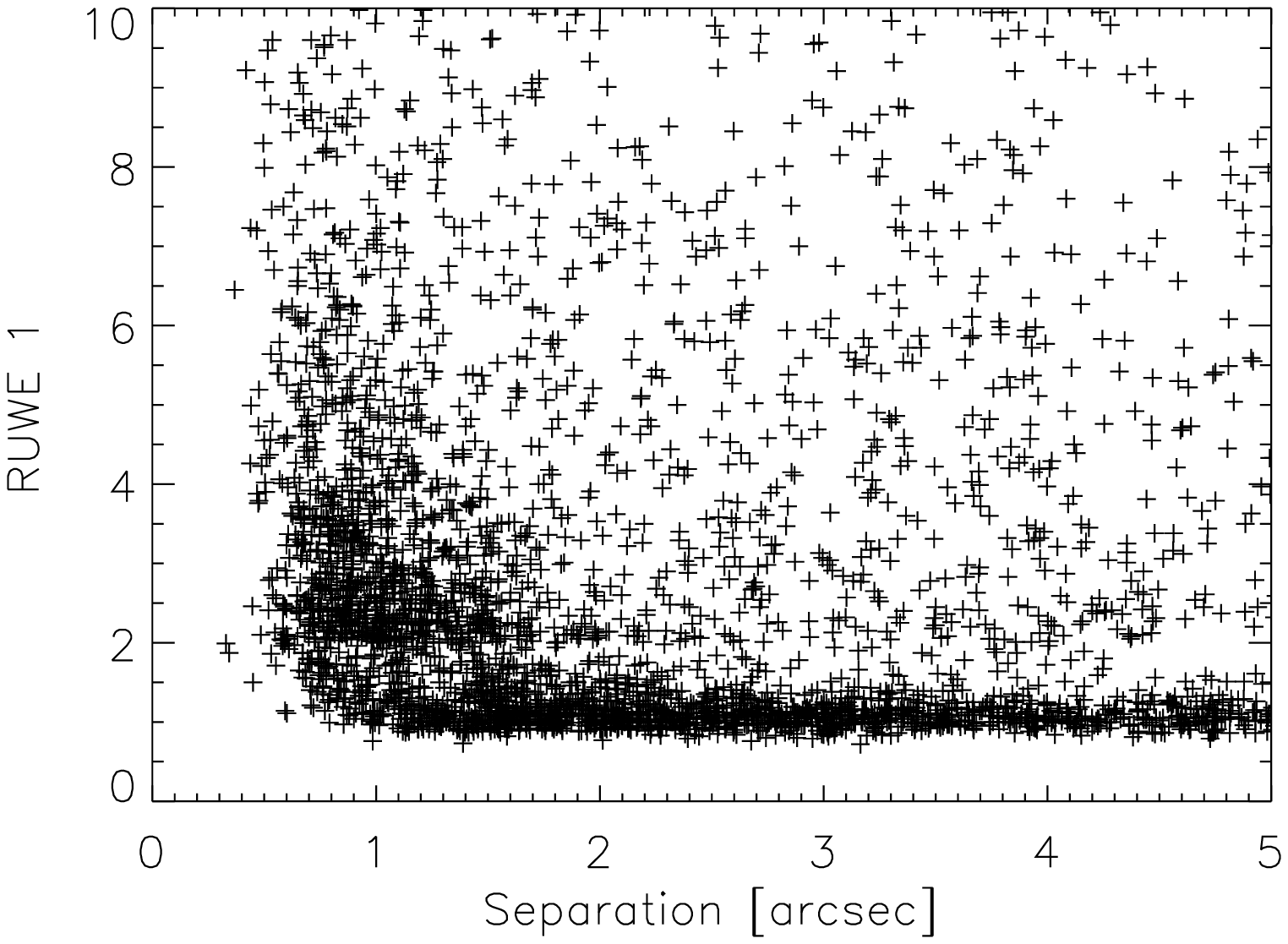}
\caption{Top:  fraction  of  double   transits  FDBL  in  the  primary
  component of resolved  Gaia binaries vs. their  separation $\rho$ in
  arcseconds.  The line is  FDBL=100*(2.5 - $\rho$)/1.5.  Bottom: RUWE
  of the primary component vs. $\rho$.
\label{fig:FDBL}
}
\end{figure}

\begin{figure}
\epsscale{1.1}
\plotone{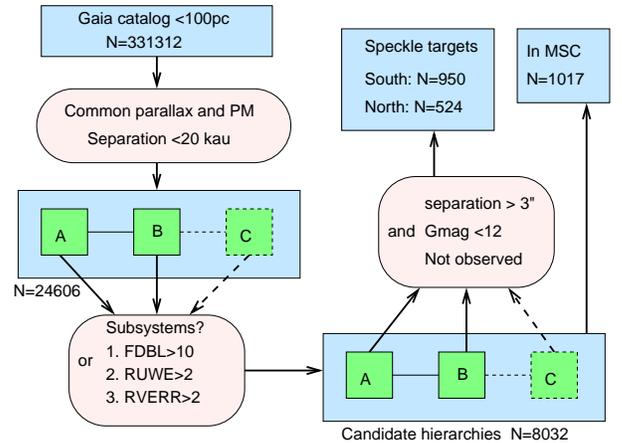}
\caption{Block diagram showing selection of candidates for
 speckle observations.
\label{fig:selection}
}
\end{figure}

The FDBL  parameter, ranging from 0  to 100, is an  even more powerful
diagnostic of a close companion than  RUWE; so far, it has received little
attention in the  literature.  However, double transits  do not always
pinpoint inner subsystems.  In a binary of $\sim$1\arcsec ~separation,
normally resolved by  Gaia as two sources, double  transits occur when
the Gaia scans are nearly parallel to  the binary.  A plot of FDBL vs.
binary  separation $\rho$  in Figure~\ref{fig:FDBL}  clearly shows  an
elevated FDBL for close pairs.  An empirical condition FDBL$> 100 (2.5
- \rho)/1.5$ inferred  from this plot separates  binaries with genuine
subsystems  from simple  binaries  with $\rho  <  2\farcs5$.  For  the
secondaries, a more  strict criterion FDBL$> 100 (3.5  - \rho)/1.5$ is
adopted, based  on a  similar plot.   However, as  shown in  the lower
panel  of Figure~\ref{fig:FDBL},  many binaries  closer than  2\farcs5
also have  an elevated  RUWE in  their primary  components, presumably
caused by  the disturbing influence  of companions on  the astrometric
measurements. A plot  of RUWE vs. the estimated speed  of orbital motion,
not  shown here,  reveals no  correlation, so  the non-linear  orbital
motion in  these binaries is unlikely  to be the cause  of an elevated
RUWE.

Figure~\ref{fig:selection}  illustrates   the  process   of  selecting
candidate hierarchies  from the  GCNS.  About 1000  systems containing
three  (or more)  related  GCNS stars  are  obvious candidates;  their
subset  has   been  studied  in   \citep{Tokovinin2022b}.   Additional
candidates are selected among wide binaries where the presence of an inner
subsystem is inferred from the  Gaia binarity flags: FDBL$>10$ (with a
higher threshold for close outer pairs  as noted above) or RUWE$>$2 or
RVERR$>$2 \kms.  Application of  these subjective criteria, adopted to
maximize the reality of subsystems, leads  to a pool of 8032 candidate
hierarchies.   In  most  cases,  the parameters of  the  inner  subsystems
(periods and mass ratios) are unknown, so this sample by itself is not
very informative for  the statistics. For this  reason some candidates
were observed at SOAR, as reported below.

The list  of 24,606 systems  (including 8032 candidate  hierarchies) is
not provided here because, given the  criteria, it can be derived from
the original GCNS.  The criteria for selecting binaries and subsystems
were  chosen subjectively,  and modified  criteria would  result in  a
different list.   The raw list has  little value, as it serves  only as a
starting  point  for  follow-up  observations  and for  additional
mining of the Gaia data.

The Multiple  Star Catalog  \citep[MSC;][]{MSC}, holds a record  of known
hierarchies  based  on  the  literature.  This  is  an  eclectic  data
collection, heavily burdened by selection  effects.  On the other hand,
the Washington Double Star Catalog \citep[WDS;][]{WDS}, holds a similarly
disparate collection of resolved  (traditionally called ``visual'') pairs,
some of which  are mere chance projections (optical  pairs).  The Gaia
candidate hierarchies  were matched  to the WDS,  and cases  where the
inner pairs in the Gaia candidates were actually resolved according to
the WDS were singled out.  Most  of those triples were already present
in the  MSC, and 150 new  ones where Gaia discovered  distant tertiary
components to the previously known visual binaries were added.  With this
increment, the MSC contained 1017 hierarchies within 100\,pc. The work
reported below has doubled this number.  However, the completeness is still
very  poor  in comparison  with  the  expected number  of  hierarchies
 in Figure~\ref{fig:trifrac}. 

About 370  hierarchical systems within  100\,pc documented in  the MSC
are not present  in our candidate list. The most  frequent classes are
tight  hierarchies  with one  or  zero  associated Gaia  sources,  and
hierarchies  where  one component  is  a  visual binary  without  Gaia
astrometry.  In a small number of cases, Gaia parallaxes are available
for both components but are strongly biased by the subsystems, so that
the two GCNS stars appear unrelated.  For example, the primary star in
01579$-$2851 has a parallax of 12.45\,mas in DR3, and 11.24\,mas after
fitting its  astrometric orbit in  the NSS; the latter  coincides with
the  parallax of  the secondary  component.  Obviously,  this pair  is
missing from the  list of candidates, which imposes  a maximum parallax
difference  of  1\,mas.  A  few  hierarchies  in  the MSC  have  outer
separations exceeding the adopted limit of 20\,kau.

\subsection{Data Organization}
\label{sec:data}

Information on binary and multiple stars in various databases is often
affected  by  confusion.   Such   attributes  as  position,  parallax,
photometry, etc.   may refer  either to the  blended light  of several
stars or  to the individual stars.   The term {\em component}  is used
here  for  referring to  the  data  on  astrometry and  photometry  of
components of multiple systems, admitting that each component may host
several stars and  that the term {\em resolved} is  fuzzy.  The notion
of component  evolves with time  as  observing  techniques improve.
Gaia provides, for the first time, resolved photometry and astrometry of
the individual   components   of   many   visual   binaries   wider   than
$\sim$1\arcsec.  At the  same time, the 2MASS  photometry and position
may still  refer to  the binary as  a whole (a  blend) because  of the
lower 2MASS resolution.  For example,  HIP~12548 is a single source in
Gaia, although  it contains  four stars  \citep{chiron8}.  In  future
Gaia data  releases it  may be  split in  two components  separated by
0\farcs4, each component hosting a close pair.

A  consistent   identification  scheme  is  implemented   in  the  MSC
\citep{MSC}. Each multiple  system has a common  10-character MSC code
based on the  J2000 coordinates of its  primary component.  Components
are designated  by letters, their  accurate coordinates for  the J2000
epoch  and other  optional identifiers  (e.g.   in the  HD or  2MASS
catalogs) are provided.  Subsystems are unions of components joined by
a comma.   In contrast,  the  WDS catalog  of  double stars  \citep{WDS}
designates systems,  rather than  components.  It  uses the  WDS codes
(10-character strings  based on  the J2000  positions), but  they may
differ from the similar MSC codes  either because a different star was
taken as the primary or because  the WDS codes are  based on inaccurate
positions (e.g. for many Luyten's wide pairs).

In  this  work, the components  of  multiple  systems that  coincide  with
individual Gaia sources are designated  by capital letters (with a few
exceptions). The Gaia own identifiers are not used because they are
not  stable, changing  between   data releases.   Instead, accurate
positions (for  J2000 or J2016  epochs) serve  to match with  Gaia and
with other  databases.  If   component  B was  resolved into  a close
pair, its members become Ba and Bb, while B refers to the blended Gaia
source.  Each system has a  unique 10-character MSC code.  This scheme
minimizes confusion and provides a direct link to the MSC.

\section{The SOAR Speckle Survey}
\label{sec:SOAR}

\subsection{Instrument, Observations, and Data Reduction}

The  high-resolution  camera,  HRCam,  is an  optical  speckle  imager
operating at SOAR since 2007  \citep{HRCam}. Over time, the instrument
got a better detector, the observing procedure has been optimized, and
the pipeline  for data processing  and calibration has  been developed
and  tuned, converting  HRCam into  a high-efficiency  survey facility
with  a typical  yield of  300  stars per  night \citep{Tok2018}.  A
survey  of binary  M-type dwarfs  \citep{Vrijmoet2022} and  imaging of
TESS exoplanet candidates \citep{Ziegler2022} demonstrate the power of
HRCam in this respect.  The  latest series of binary-star measurements
and an overview of the  ongoing HRCam observing programs are published
in \citet{Tokovinin2022a}.

Observations  of candidate  hierarchies from  the GCNS  were conducted
since 2021 October  as a filler among other  observing programs, using
also     some      engineering     time.      As      indicated     in
Figure~\ref{fig:selection},  speckle targets  were  selected from  the
pool of  8032 candidates using  additional criteria, namely $G  > 12$
mag (fainter stars  require excellent seeing conditions),  and $\rho >
3\arcsec$  (to  avoid  potentially  false candidates  caused  by  wide
companions).  Only the  RUWE and FDBL flags were  considered, and only
previously unobserved stars with a decl. south of $+20\degr$ ~were
placed on the SOAR program, resulting in 950 targets.  A complementary
list of 524  northern targets was produced. No  speckle instruments in
the Northern  Hemisphere matching HRCam in  productivity are available
to make the northern extension of this survey a practical undertaking:
it would require $\sim$10 nights at a 4 m telescope.

The main  survey started in  2022 January.   It was preceded  by trial
observations of  candidate hierarchies  selected by  various criteria.
In the last months of 2022, the observing program has been extended by
adding candidate  pairs with  separations from 1\arcsec  ~to 3\arcsec.
All  results are  reported  here  jointly.  The  $G  <  12$ mag  limit
corresponds to  $ M >  0.7$\msun at 100\,pc.  The median mass  of the
observed stars is 0.82 \msun, and  80\% are comprised between 0.60 and
1.23 \msun.  The  median mass of the resolved stars  (estimated in the
same manner from their combined absolute magnitude) is 0.80 \msun, and
80\%  are between  0.59 and  1.13  \msun. Thus,  the resolved  speckle
targets  are,  on  average,   slightly  fainter  than  the  unresolved
ones.  The likely  explanation of  this difference  is a  better prior
coverage of bright stars (previously resolved pairs were not placed on
the program).

In each  HRCam observation, two  image cubes of  200$\times$200 pixels
and 400  frames are taken  with an exposure  time of 25\,ms  (8\,s per
cube) and a pixel scale of 15\,mas.  The observations were made in the
$I$ filter  (824/170nm) to maximize  the flux  from red stars  and the
detectability of  faint red companions.  The  classical resolution limit
set by  diffraction is 40\,mas,  but closer pairs of  near-equal stars
were detected  down to  30\,mas separation from  the asymmetry  of the
speckle   power  spectrum;   measurements  of   their  positions   are
inaccurate.   Some  close  pairs   were  reobserved  to  confirm  the
detections  and to  follow their  expected fast  orbital motion.   The
approximate  detection   limits  (resolution  and   maximum  magnitude
difference vs.  separation) are determined by the speckle pipeline for
each  observation. They  depend on  the seeing  conditions and  on the
target  brightness.   The  contrast  limit  at  0\farcs15  separation,
$\Delta I_{0.15}$,  is increased here by  0.5 mag with respect  to the
original conservative  estimates delivered  by the pipeline  to reflect better
parameters of the resolved pairs.  The astrometric calibration
is common to all HRCam programs.

\begin{deluxetable*}{l l  c c  cccc c ccc c l}
\tabletypesize{\scriptsize}     
\tablecaption{Results of the SOAR Speckle Survey (fragment)
\label{tab:speckletable} }  
\tablewidth{0pt}                                   
\tablehead{     
\colhead{MSC} &  
\colhead{Comp.} &  
\colhead{R.A.} &  
\colhead{Dec.} &  
\colhead{Date} &  
\colhead{$\theta$} &  
\colhead{$\rho$} &  
\colhead{$\Delta I$} &  
\colhead{Flag} &  
\colhead{$\rho_{\rm min}$} &  
\colhead{$\Delta I_{0.15}$} &  
\colhead{$\Delta I_1$} &  
\colhead{NSS} & 
\colhead{System} \\
(J2000) &   &
\colhead{(deg)} &  
\colhead{(deg)} &  
\colhead{(JY-2000)} &  
\colhead{(deg)} &  
\colhead{($''$)} &  
\colhead{(mag)} &  &
\colhead{($''$)} &  
\colhead{(mag)} &  
\colhead{(mag)} & &  
}
\startdata
00025$+$0440 &AC &  0.613899 &  4.668529   &21.8909& 196.7 & 1.0596 &  0.2& *       &0.052 &2.44 &3.98 &---  & AC \\
00026$-$2814 &A  &  0.661087 &$-$28.236648 &22.4418&   0.0 & 0.0000 &  0.0& \ldots  &0.057 &2.24 &3.37 &AORB & UR \\
00042$-$1008 &A  &  1.051757 &$-$10.141044 &22.4419&   0.0 & 0.0000 &  0.0& \ldots  &0.044 &2.44 &4.40 &---  & UR\\
00049$-$1811 &A  &  1.237158 &$-$18.178732 &22.4419&   0.0 & 0.0000 &  0.0& \ldots  &0.043 &2.32 &4.46 &---  & UR \\
00092$-$0408 &A  &  2.306404 & $-$4.133919 &21.7542& 351.5 & 0.0889 &  0.3& q       &0.051 &2.10 &3.79 &---  & KPP2684Aa,Ab \\
00092$-$0408 &A  &  2.306404 & $-$4.133919 &22.4446&   0.0 & 0.0000 &  0.0& \ldots  &0.050 &2.70 &3.71 &---  & KPP2684Aa,Ab \\
00092$-$0408 &A  &  2.306404 & $-$4.133919 &22.6823&   0.0 & 0.0000 &  0.0& \ldots  &0.053 &2.74 &3.73 &---  & KPP2684Aa,Ab \\
00100$-$5358 &A  &  2.491435 &$-$53.958769 &22.4447&   0.0 & 0.0000 &  0.0&\ldots   &0.046 &2.94 &4.35 &ASB1 & UR \\
00111$-$0008 &B  &  2.725629 & $-$0.111976 &22.4447& 305.8 & 0.3546 &  2.5& q       &0.051 &2.65 &3.81 &---  & Ba,Bb \\
00119$-$3533 &A  &  2.962362 &$-$35.546968 &22.8452& 125.4 & 0.0974 &  1.9& q       &0.043 &2.72 &3.47 &---  & Aa,Ab \\
00119$-$3533 &A  &  2.962362 &$-$35.546968 &23.0062& 126.0 & 0.0963 &  1.9& q       &0.041 &3.07 &5.76 &---  & Aa,Ab \\
\enddata
\end{deluxetable*}

\begin{figure}
\epsscale{1.1}
\plotone{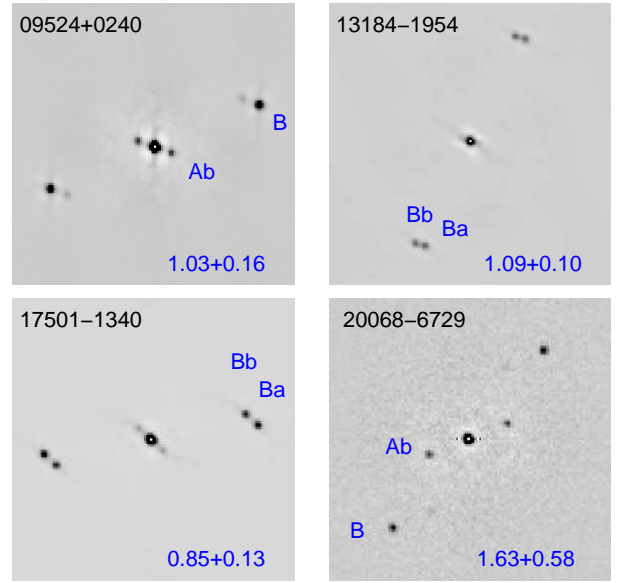}
\caption{New triple  stars where  the wide  companions are  present in
  Gaia DR3 and the inner subsystems are resolved by SOAR.  The speckle
  ACFs are displayed  with an arbitrary negative  stretch to highlight
  the  companions.    Each  panel  has   an  MSC  label.    The  peaks
  corresponding to the companions are  marked, and the outer and inner
  separations in arcseconds are indicated.
\label{fig:oldtriples}
}
\end{figure}

\begin{figure}
\epsscale{1.1}
\plotone{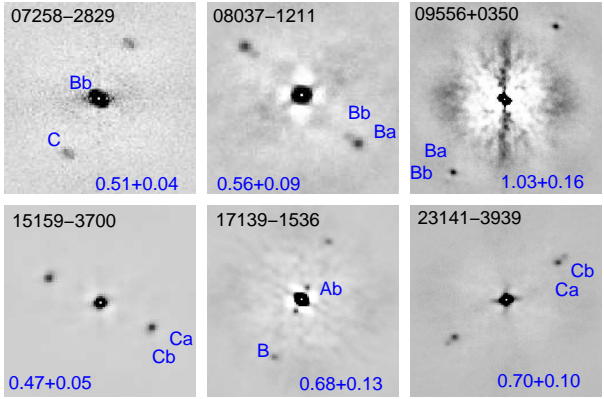}
\caption{New compact  triplets discovered by the  SOAR speckle imaging
  of Gaia candidates. See the caption to Figure~\ref{fig:oldtriples}.
\label{fig:newtriples}
}
\end{figure}

\subsection{Results} 

Table~\ref{tab:speckletable}, published fully electronically, presents
the results of this survey. Its  first columns contain the MSC code of
the system (similar  to the WDS code, but not  always coincident), the
component's identifier,  and its  accurate equatorial  coordinates for
the  J2000  epoch from  the  GCNS.  This information  should  uniquely
identify each observed target. When  the measurement involves two Gaia
sources, the  component identifier has two  characters.  The following
columns  contain  Julian  year  of  the  observation,  position  angle
$\theta$, separation $\rho$, and  magnitude difference $\Delta I$; for
unresolved targets all these numbers are zero.  Note that 00092$-$0408
A has  been resolved in 2021.75  at 0\farcs089, but unresolved  on two
occasions in 2022.  An optional  flag after $\Delta I$ indicates cases
where the magnitude difference is determined from the average image of
a  wide  pair (*),  when  the  quadrant  is defined  without  180\degr
~ambiguity (q), the data are noisy or below the diffraction limit (:),
and a few observations of close pairs in the Str\"omgren $y$ band (y).
The three following columns contain  the detection limits: the minimum
separation $\rho_{\rm min}$, and  the maximum magnitude differences at
0\farcs15  and 1\arcsec  ~separations, $\Delta  I_{0.15}$ and  $\Delta
I_1$, respectively.  The next column gives a code of the NSS solution,
if  present (see  Section~\ref{sec:orb}) or  --- otherwise.   The last
column  contains  the  WDS  discoverer  codes  of  the  systems  where
appropriate (e.g.  KPP2684Aa,Ab  for the resolved primary  star of the
3\farcs6 Gaia pair  named KPP2684 in the  WDS), otherwise designations
like Ba,Bb for the newly resolved pairs, UR for unresolved sources, AB
for Gaia pairs, or Aa,Ab and Aa,B for resolved triples.

Table~\ref{tab:speckletable}  contains 1384  entries corresponding  to
1243 unique targets (either single  components or pairs); 1058 of them
have  single-letter component  identifiers,  503 of  which (48\%)  are
resolved.  There  are 47 observed targets  fainter than $G =  12$ mag,
eight of those are fainter than $G = 14$ mag, and the faintest one has
$G=17.3$  mag.  Faint  targets with  strong indications  of subsystems
were observed  as a complement to  the main survey, like  pairs closer
than 3\arcsec.
 
The standard  SOAR speckle  pipeline \citep{Tok2018}  delivers speckle
power  spectra   and  image  autocorrelation  functions   (ACFs).   To
illustrate the nature of these data and to highlight some discoveries,
Figures~\ref{fig:oldtriples} and \ref{fig:newtriples} show the ACFs of
targets with three  stars in the HRCam field.  The  wide components in
Figure~\ref{fig:oldtriples} are  actually found in Gaia  DR3, but when
they are  themselves close  pairs, Gaia does  not have  parallaxes, so
these components are missed in the GCNS and, consequently, in our list
of candidate hierarchies.  Nevertheless,  these stars have additional,
wider companions in the GCNS, so these systems are at least quadruple.
In the  20068$-$6729, the outer component  C is only at  4\farcs15 and
96\fdg4 from A, so the whole quadruple (including the newly discovered
component  Ab)  has small  ratios  between  separations and  could  be
dynamically  unstable.   At a  parallax  of  11.5\,mas, the  estimated
periods in this system range from 300  yr to 5 kyr.  The elevated RUWE
of stars  A and  B (5.6  and 2.4, respectively)  is likely  caused by 
light from the new star Ab.

Most  inner  subsystems  resolved  by SOAR  are  binaries,  but  some,
unexpectedly, contain three stars. Six ACFs of such triplets are shown
in  Figure~\ref{fig:newtriples};  they  have outer  separations  below
1\arcsec  ~and inner  separations  near the  resolution limit.   These
compact triplets  have only one component  in the GCNS, but  there are
other, more  distant Gaia  components, making  these systems  at least
quadruple.   07258$-$2829 and  09556+0350 are  actually quintuples
because their  distant components are  also resolved at SOAR  as close
pairs.  The inner pair Ba,Bb in 09556+0350 is barely seen because star
Bb, 6.7  mag fainter than  A, is  below the formal  contrast detection
limit.

\begin{figure}
\epsscale{1.1}
\plotone{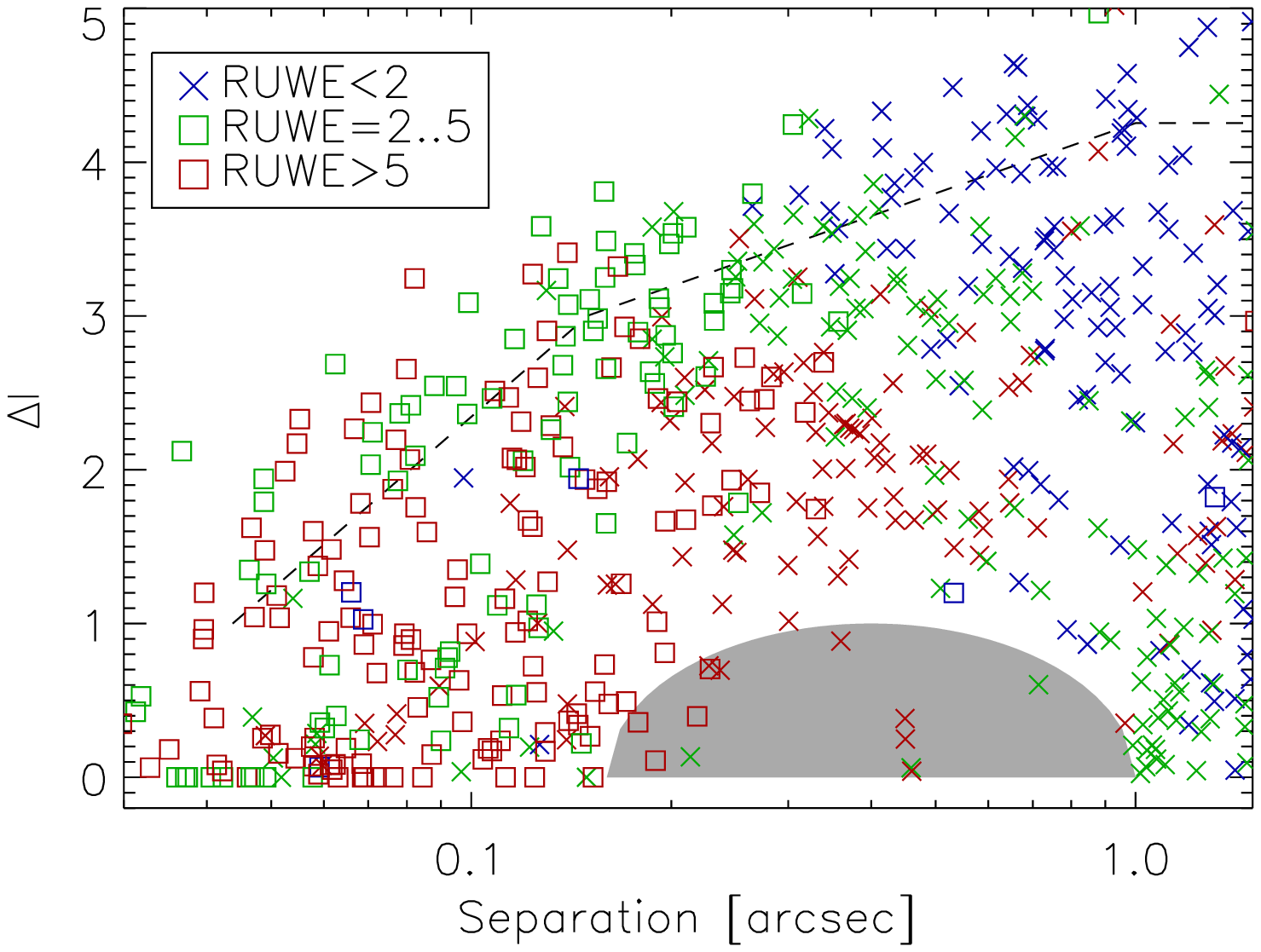}
\plotone{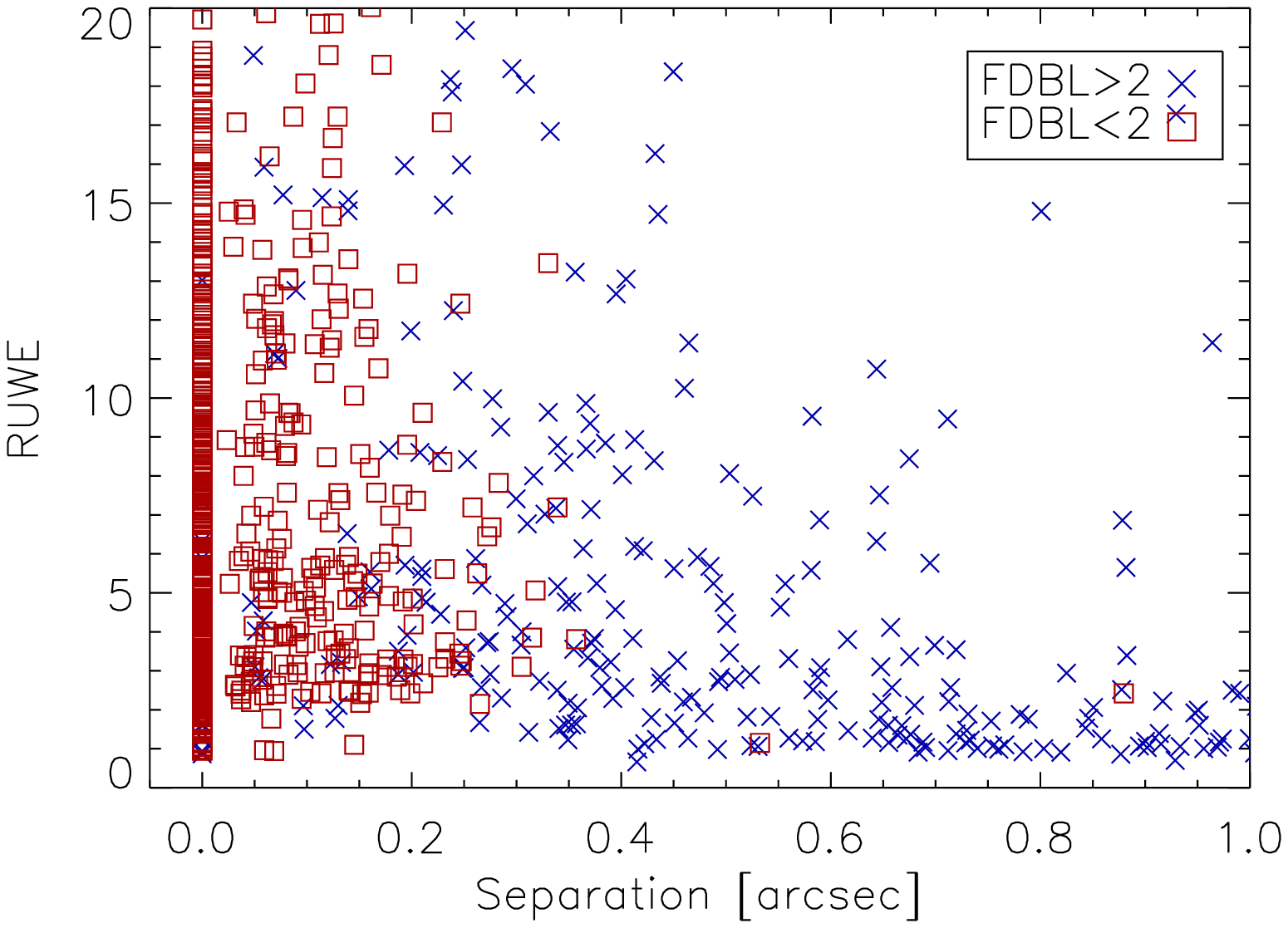}
\caption{Comparison between  SOAR resolutions of  candidate subsystems
  and Gaia binarity  flags.  Top: magnitude difference  $\Delta I$ vs.
  separation   for   resolved   pairs.   Crosses   indicate   FDBL$>$2
  (effectively resolved  by Gaia), and the squares mark  FDBL$<$2.  The colors
  show RUWE: $<$2 (blue), between 2 and 5 (green), and $>$5 (red). The
  gray shade indicates the Gaia avoidance of binaries, the dashed line
  is  the  median detection  limit.  
   Bottom: RUWE vs. angular separation.
  The blue crosses and red squares distinguish targets by FDBL.
\label{fig:speckle}
}
\end{figure}

\subsection{SOAR Resolutions vs. Gaia Binarity Flags}

The  top panel  of Figure~\ref{fig:speckle}  is a  separation--contrast
plot for  the resolved subsystems.  At the same time, it compares
with the  Gaia binarity flags  (crosses with FDBL$>$2  are effectively
resolved by Gaia, squares have single transits), while colors code the
RUWE.   Essentially all  pairs wider  than 0\farcs2  (and some  closer
ones)  are  resolved  by  double transits.   The  widest  and  dimmest
companions have small RUWEs (blue crosses) and are not revealed by this
criterion. On the other hand, brighter  companions with $\Delta I < 2$
mag  at separations  of  $\sim$0\farcs5 have  red  and green  symbols,
indicating that an  elevated RUWE was likely caused  by the perturbing
light of those companions rather than  by the slow orbital motion with
estimated  periods  on the  order  of  centuries.  Pairs  closer  than
0\farcs1   are  detected   only   by  RUWE.    The   lower  panel   of
Figure~\ref{fig:speckle} gives  a complementary view of  the interplay
between  $\rho$,  RUWE,  and  FDBL.   One  can  note  that  {\em  all}
subsystems identified by the FDBL  flag (blue crosses) are resolved at
SOAR.   In contrast,  a large  part of  subsystems with  elevated RUWEs
remained  unresolved (zero  separation)  either because  they are  too
close  or  because  the  companions  are  too  faint.   Although  some
correlation between parameters of the resolved pairs $(\rho, \Delta I)$
and  the  Gaia  binarity  flags   is  incontestable,  this  is  not  a
deterministic relation because additional  factors (e.g.  the number of
Gaia transits) influence the Gaia binarity indicators.

Figure~\ref{fig:speckle} (top) shows a deficit of pairs with magnitude
differences below 1  mag and separations above  0\farcs2.  This ``Gaia
hole'' is caused by missing parallaxes of close binaries, as mentioned
above, so these stars are not  present in the GCNS. The lower envelope
gives  approximate  limits of  the  hole  in  the $(\rho,  \Delta  I)$
space.  It can  be  described  crudely by  a  semi-circle centered  at
0\farcs4 with  a logarithmic width  of 2.5$\times$  and a height  of 1
mag:
\begin{equation}
[ \log(\rho/0\farcs4)/\log(2.5)]^2 + (\Delta I)^2 < 1
\label{eq:hole}
\end{equation}
(see  the gray  shading  in  Figure~\ref{fig:speckle}).  However,  the
limits are  fuzzy, apparently depending  on the Gaia scanning  law and
source location  (some binaries  may be  lucky in  getting parallaxes,
while  other  binaries with  similar  parameters  are not).  Note  the
crosses  near $(0\farcs4,  0)$ ---  binaries where  Gaia DR3  measured
parallaxes  of  one or  both  components  with comparable  magnitudes.
Knowing the shape of the Gaia  hole and the binary statistics, one can
estimate the number of stars missing from the GCNS.

The resolution of  inner subsystems at  SOAR enables estimation  of masses
and mass  ratios (from  absolute magnitudes, using  standard relations
for  main-sequence  stars), as well as  periods  (assuming  that  projected
separation is statistically representative of the semimajor axis). The
methods  are  explained  in  the   MSC  paper  \citep{MSC}.   All  new
hierarchies with resolved subsystems are added to the MSC, which holds
additional parameters  such as estimated masses,  periods, astrometry,
etc.   This  information   is  not  duplicated  here,   only  the  new
observations are reported in Table~\ref{tab:speckletable}. The updated
MSC is publicly available through Vizier (catalog J/ApJS/235/6) and at
\url{http://www.ctio.noirlab.edu/~atokovin/stars/}.

\section{Hierarchies with Inner Gaia Orbits}
\label{sec:orb}

Another input to  the statistics of nearby hierarchies  is derived from
the      orbital      solutions      in      the      NSS      catalog
\citep{Arenou2022,Pourbaix2022}.  The NSS  information is presented in
17 tables, separately  for each solution type.  Here, only  the eight most
frequent  types  are  used,  ignoring the  rest  (eclipsing  binaries,
circular orbits,  etc.). The data were recovered from  the Vizier
catalog  I/357 \citep{NSS-Vizier}  and ingested  into IDL  structures.
Note  that the  SB1 and  SB2  tables in  the  NSS do  not contain  any
astrometric  information.    The  Gaia  astrometry   of  spectroscopic
binaries was recovered  from the main DR3 catalog, linked  to the NSS
by the Gaia identifiers.

\begin{deluxetable}{l l l l }
\tabletypesize{\scriptsize}     
\tablecaption{NSS Solutions
\label{tab:nss} }  
\tablewidth{0pt}                                   
\tablehead{     
\colhead{Code} & 
\colhead{Solution} & 
\colhead{$N_{\rm mult}$}  &
\colhead{$N_{\rm GCNS}$}
}
\startdata
AORB & Orbital       &   251 & 2355 \\
ASB1 & AstroSpectroSB1 & 144 & 1400 \\
SB1  & SB1 &           231 & 1388 \\
SB2  & SB2 &           79  & 311 \\ 
A7   & Acceleration7 & 331 & 2398 \\ 
A9   & Acceleration9 & 221 & 1970 \\
RV1  & FirstDegreeTrendSB1 & 53 & 277 \\ 
RV2  & SecondDegreeTrendSB1 & 55 & 289 \\
All  & \ldots               & 1365 & 10,388 \\ 
\enddata
\end{deluxetable}

\begin{deluxetable}{l l l c c c l}
\tabletypesize{\scriptsize}     
\tablecaption{Resolved Subsystems with NSS Orbits
\label{tab:nssspeckle} }  
\tablewidth{0pt}                                   
\tablehead{     
\colhead{MSC} &  
\colhead{Comp.} & 
\colhead{Sol.} & 
\colhead{$\rho$} & 
\colhead{$\Delta I$} & 
\colhead{$P$}  & 
\colhead{Comment} \\
  &  & &
\colhead{($''$)} &
\colhead{(mag)} &
\colhead{(d)} &
}
\startdata
01242$-$2157 &A& ASB1   & 0.057  &  0.2  & 1734 & Blended \\ 
04097$-$5256 &A& AORB   & 0.176  &  3.4  & 121.9  & Suspect \\ 
05130$-$8125 &A& SB1    & 0.040  &  1.1  & 1192  & Blended \\ 
05574$-$2458 &A& SB1    & 0.467  &  3.1  & 378  & Suspect \\ 
07058$-$5849 &A& SB1    & 0.052  &  2.0  & 1310 & OK \\
07530$-$0201 &A& SB1    & 0.199  &  2.3  & 1.5  & Wrong \\ 
07543+0232   &A& SB1    & 0.364  &  2.3  & 36.6  & Quintuple? \\ 
09336$-$2752 &A& SB2    & 0.139  &  2.9  & 37.5   & Quadruple? \\ 
10356$-$4715 &A& ASB1   & 0.026  &  0.0  & 547.4  & Blended \\
12022$-$4844 &A& SB1    & 0.095  &  3.1  & 52.0  & Quadruple? \\ 
15229$-$6242 &B& AORB   & 0.047  &  0.7  & 1421 & OK, moves \\ 
15397$-$4956 &A& SB1    & 0.046  &  0.8  & 44.8  & Suspect \\ 
16412+0349   &A& SB1    & 0.057  &  1.3  & 1034 & Blended  \\ 
19221$-$0444 &B& AORB   & 0.086  &  1.6  & 2624 & OK \\ %
19369$-$6949 &A& ASB1   & 0.023  &  0.0  & 735  & Blended \\
20147$-$7252 &A& SB1    & 0.253  &  3.5  & 305.9 & Suspect \\ 
21320$-$0129 &B& ASB1   & 0.030  &  0.7  & 1056 & Blended \\
21460$-$5233 &A& SB1    & 0.088  &  2.5  & 28.6  & Suspect \\ 
22170+1824   &B& SB1    & 0.366  &  2.0  & 11.9  & Quadruple? \\ 
22377$-$0210 &A& SB1    & 0.049  &  1.7  & 7.0   & Quadruple \\ 
\enddata
\end{deluxetable}

Table~\ref{tab:nss} gives  the short  codes adopted  here for  the NSS
solutions, their  official self-explanatory names, the  number $N_{\rm
  mult}$ of such solutions found  among wide binaries, and their total
number  $N_{\rm GCNS}$  in the  GCNS.   The first  four types  provide
orbital elements and thus are  relevant for the statistics.  Those 705
subsystems were matched  to the MSC and added to  it, if missing.  The
other half  of the solutions  (660) give only accelerations  or radial
velocity (RV)  trends; they  do not constrain  the inner  mass ratios,
while the  periods likely exceed 1000  days.  The total number  of NSS
solutions for  the GCNS objects  is 10,388,  or 3.1\%.  The  number of
solutions for the 50,243 members of wide pairs or triples is 1365, or
2.7\%.   By  construction,  the  NSS  catalog  tried  to  avoid  close
companions, explaining its slightly lower  rate of solutions  for stars
belonging to wide binaries.

The total numbers  of stars with RUWE$>$2  (candidates for astrometric
orbits) are  40,336 and 6995  in the full GCNS  and in our  list of
50,243 stars, respectively.  The numbers of astrometric (AORB and ASB1)
orbits in  Table~\ref{tab:nss}, 395 and  3755, are much  smaller, only
9.3\% and  5.6\% of stars  with RUWE$>$2. The speckle  survey suggests
that half  of the  RUWE-selected candidates can  be resolved,  so that
their  long periods  are  not  yet covered  by  the  NSS.  Still,  the
estimated  $\sim$19\%  completeness  of  astrometric  orbits  for  the
remaining half is quite low. 

There are 318  matches between Nthe SS solutions and  speckle targets, and
36  of  those  are  resolved  by SOAR  at separations below 1\arcsec.   The
resolution  rate of  11\% is  substantially  lower than  for the  whole
speckle  survey  (48\%).  Only  20  resolved  subsystems have  orbital
elements   in  the   NSS.    However,  the   detailed  comparison   in
Table~\ref{tab:nssspeckle}  casts   doubts  on  some   orbits.   Close
companions  perturb  Gaia  astrometry  and the  RVs  measured  by  its
slitless spectrograph, leading to suspicious orbits.  For example, the
orbit of  07530$-$0201 A  with $P=1.5$  days and  amplitude $K_1  = 1$
\kms, if  true, would  imply an unlikely  substellar companion  in the
brown dwarf desert  regime, so the period  seems spurious.  Suspicious
orbits with periods close to a year or its harmonics could result from
the companion-induced  effects that vary  in a regular  way throughout
the year,  following the Gaia scanning  direction.  Some spectroscopic
orbits with periods on the order of a month or shorter could be real,
indicating that the companion resolved by SOAR (with estimated periods
of a few years or decades) orbits an inner spectroscopic binary, while
the  resolved Gaia  companion  is  on a  still  wider  outer orbit  (a
quadruple  of 3+1  hierarchy).  However,  the possibility  that some  of
those spectroscopic orbits are wrong  still remains, and  follow-up
RV monitoring is needed for their verification.

In  eight   resolved  subsystems   the  periods  estimated   from  the
separations approximately match the NSS orbital periods.  However, the
amplitude of  the RV variation  or the astrometric semimajor  axis are
often reduced  by blending of comparable-brightness  stars, as follows
from the measured magnitude differences.  The mass ratios derived from
the  blended NSS  orbits are  in  fact lower  limits. Several  speckle
measurements of  10356$-$4715 and  19369$-$6949 taken during  one year
show  rapid motion compatible with their 2 yr orbits.

The uniform and impersonal coverage of Gaia orbits offers a definitive
advantage  for  statistical  studies.  However,  the  automatic  orbit
calculation and the cadence imposed by the Gaia scanning law lead to a
non-negligible  fraction of  wrong orbits,  despite efforts  to remove
them by the NSS creators.


\section{Discussion}
\label{sec:disc}

\begin{figure}
\epsscale{1.1}
\plotone{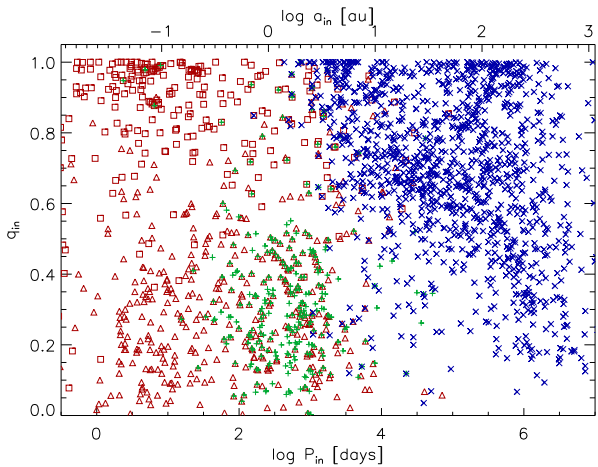}
\caption{Periods  and  mass  ratios   of  inner  subsystems  in  known
  hierarchies within 100 pc. The SB1 and SB2 orbits are plotted by the
  red squares and triangles, respectively, astrometric orbits by green
  pluses, and resolved  subsystems by blue crosses.  The mass ratios
  of SB1s and  astrometric binaries are lower limits.   The upper axis
  gives orbital separations for a mass sum of 2 \msun.
\label{fig:psqs}
}
\end{figure}

\begin{figure}
\epsscale{1.1}
\plotone{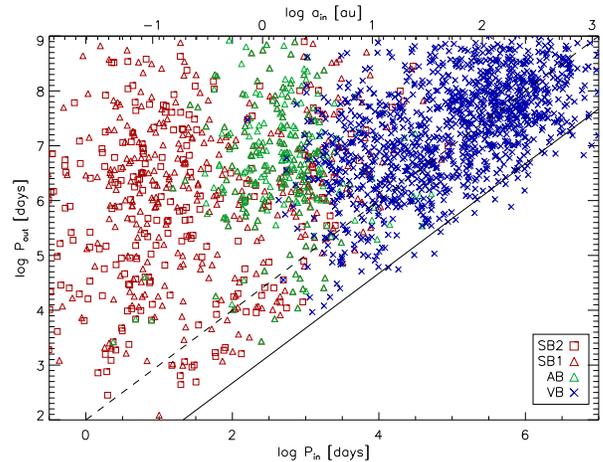}
\caption{Periods  in  the  inner  and  outer  subsystems  at  adjacent
  hierarchical levels. The symbols correspond to the detection methods
  of inner subsystems, as in Figure~\ref{fig:psqs}. The solid and dash
  lines indicate the period ratios of 4.7 and 100, respectively.
\label{fig:plps}
}
\end{figure}

After the  addition of  the newly  resolved subsystems  and subsystems
with NSS orbits, the MSC contains 2,754 hierarchies within 100 pc with
estimated periods.   For the following  discussion, I select  from the
MSC 2,208 systems  within 100\,pc with primary masses from  0.5 to 1.5
\msun (excluding 77  systems containing known white  dwarfs).  Most of
them  are  triples,  while  systems  of four  or  more  stars  can  be
decomposed  into elementary  triples. Figure~\ref{fig:psqs}  plots the
inner  periods  and   mass  ratios  for  this   sample.   The  symbols
distinguish the observing  techniques and clearly separate  the subsystems
into groups.   The upper right corner  of the plot is  occupied by the
1,280  resolved  subsystems    (blue  crosses),  including  those
studied  here. Their  lowest  mass  ratios depend  on  the period  (or
separation) owing to  the limitation of the  observing method (speckle
and Gaia  resolutions).  The  243 double-lined  spectroscopic binaries
 (red squares)  occupy the upper-left corner,  as expected.  Most
of the  444 single-lined spectroscopic   (red triangles)  and 273
astrometric (green pluses) binaries  have periods shorter than 3
yr  (duration of the  Gaia DR3  observations), and  their mass  ratios are
lower limits owing to blending.   Orbital inclination also reduces the
mass ratios  of SB1s,  but this  is a smaller  effect compared  to the
blending.  Please, keep in mind that some NSS orbits are false.

The gaps between the three  islands of points in Figure~\ref{fig:psqs}
are almost certainly  caused by the observing  techniques, rather than
by a real dichotomy of the underlying population.  Overlapping symbols
correspond  to simultaneous  detections  by several  methods, but  the
number of  such overlaps  is quite  modest. The  density of  points at
short  periods of $P_{\rm  in} <  1000$  days is  less than  in the  area
covered by the speckle detections,  reflecting the small percentage of
NSS  orbits mentioned  above.   The plot  gives a  rough  idea of  the
coverage of the  parameter space and of  the remaining incompleteness.
For example, many subsystems with $a_{\rm in} \sim 10$ au and $q_{\rm in} <
0.6$, below  the speckle detection  limit, are  yet to be  resolved by
high-contrast  imaging.  Their orbital  periods  are  longer than  the
duration of the Gaia mission.

Let  us  focus on  the  upper-right corner  of  Figure~\ref{fig:psqs}
($P_{\rm in} > 10^4$ days, $q_{\rm  in} > 0.6$), where the detection of
subsystems by imaging  is uniform.  The observed  distribution is thus
representative of the real distribution of subsystems, and it has some
interesting features  worthy of  comment.  The concentration  of points
near  $q_{\rm in}  \approx 1$  is a  well-known manifestation  of twin
binaries  (it  is   even  more  prominent  at   shorter  periods).   A
statistically  significant  excess  of   wide twin  binaries  with
separations up to $10^3$ au  has been detected by \citet{ElBadry2019}.
The recent  study by  \citet{Hwang2022a}  shows  that wide  twins  with
separations from  400 to  $10^3$ au  have extremely  eccentric orbits,
suggesting that they were formed as tighter 10--100 au pairs and later
ejected  to  wide  and   eccentric  orbits,  presumably  by  dynamical
interactions in unstable triples. In  the light of this discovery, the
abrupt decrease in the frequency of inner subsystems (including twins)
at  separations  above  $\sim$300  au,  seen  in  Figure~\ref{fig:psqs},
appears  natural  (wider  subsystems form  rarely).   This  separation
corresponds to 3\arcsec ~at 100\,pc distance, beyond the Gaia hole that
ends at  1\arcsec.  So, the paucity  of wider inner subsystems  is not
caused by  observational selection.

Detailed examination of these data  reveals that most inner pairs with
$q_{\rm in} \sim 1$ and separations from 10 to 100 au are missing from
our list of  candidate hierarchies because they fall in  the Gaia hole
(there are fewer points in this  area of the plot).  These hierarchies
are known owing to historic ground-based efforts.  One can also note a
slight deficiency  of wide inner  subsystems with $q_{\rm  in} \approx
0.8$.  If the  significance of this local minimum  in the distribution
of  $q_{\rm  in}$   is  confirmed  and  proven  not   to  result  from
observational  biases, its  explanation  will  present an  interesting
challenge to the theory of multiple-star formation.

Figure~\ref{fig:plps} is a standard plot comparing  inner and outer
periods in   nearby hierarchies, with symbols  coding the detection
techniques of  inner subsystems in  the same  way as above.   The Gaia
resolution limit of $\sim$1\arcsec ~corresponds  to 100 au or a period
of $10^{5.5}$ days.  The outer  periods of the Gaia candidates studied
here are located  above this line, and one notes  a reduced density of
points  below it,  at shorter outer  periods. This  region of  the parameter
space  where Gaia  does not  discover new  hierarchies suffers  from 
a larger  incompleteness. Systematic  discovery and  characterization of
compact hierarchies  with outer  separations below  100 au  remains an
outstanding observational challenge. 

Some  points  in  Figure~\ref{fig:plps}  for  resolved  triples  (blue
crosses) fall below the  dynamical stability limit $P_{\rm out}/P_{\rm
  in} >  4.7$, depicted by the  solid line.  The periods  are estimated
only  crudely from  projected  separations,  explaining this  apparent
contradiction.   Nevertheless,  there  is   a  substantial  number  of
marginally  stable or  even unstable  hierarchies.  Marginally  stable
hierarchies with short periods $P_{\rm  in} < 100$ days are especially
interesting  because  their  non-Keplerian   motion  can  be  directly
observed \citep[see  an example  in][]{Borkovits2019}.  Unfortunately,
Gaia is  of little  help for  the study  of these  fascinating objects
because it was not designed for such work.

\begin{figure}
\epsscale{1.1}
\plotone{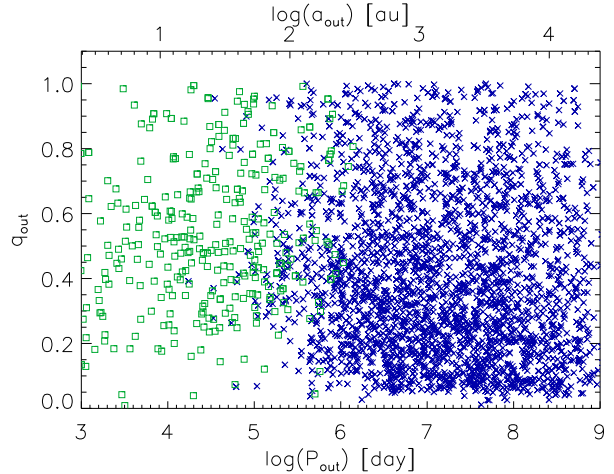}
\caption{Periods  and mass  ratios in  the outer  subsystems of  known
  hierarchies within 100  pc. The blue crosses denote  wide (common PM)
    outer  pairs, and the green  squares  correspond  to other  (mostly
  visual and speckle) discovery methods.
\label{fig:plql}
}
\end{figure}

Figure~\ref{fig:plql}  presents  outer  periods and  mass  ratios  for
hierarchies within 100\,pc. Wide  tertiary companions with $P_{\rm out}
> 10^6$  days have  separations  above 300\,au  (3\arcsec ~at a 100\,pc
distance), therefore their detection by Gaia is quite complete, unlike
closer  companions  detected  by  high-resolution  imaging.   A  mild
decrease of the outer mass ratio  with increasing outer period is thus
a  real feature  of   nearby  hierarchies, rather  than a  selection
effect. The  median $q_{\rm  out}$ is 0.39  for $P_{\rm  out}$ between
$10^6$ and $10^7$ days; it decreases to  0.36 and 0.34 in the next two
decades of  outer periods.  At outer periods below $10^{5.5}$ days,
covered  mostly by  imaging techniques,  large outer  mass ratios  are
rare; the  points group near $q_{\rm  out} \sim 0.5$, as  expected for
triples composed of three similar-mass  stars that are discovered more
readily.

In  this  work, hierarchies  are  identified  by searching  for  inner
subsystems in  wide binaries.  \citet{Hwang2023} did  the opposite by
looking  for  wide tertiary  companions  to  close binaries  with  NSS
orbital solutions.  The wide-binary catalog of \citet{ElBadry2021} was
used,  and  an  outer  separation  range between  1  and  10  kau  was
considered.  The  sample was restricted to the main-sequence stars within
500\,pc with  masses from 0.8 to  1.4 \msun.  Such field  stars have a
5.35\% fraction of wide companions, but  this fraction was found to be
larger  by   2.28$\pm$0.10  times   for  eclipsing  binaries   and  by
1.33$\pm$0.05  times  for  SBs.  The enhanced  frequency  of  tertiary
companions to close solar-type binaries discovered in  Gaia data by
Hwang confirms  the earlier results \citep{Tok2006,MSC}  and indicates
that the formation of close and wide subsystems is somehow related.

For astrometric binaries,  the frequency of tertiaries  found by Hwang
was  only a 0.65$\pm$0.03  fraction  of their  frequency  in the  field.
However, the presence of  astrometric subsystems biases parallaxes and
PMs, so  many  wide binaries  with astrometric  subsystems
appear  unbound  and get  excluded  from  the El-Badry's  catalog,  which
imposes the boundness condition. This pitfall is avoided here by using
a relaxed criterion for  wide-binary selection. Our sample  of wide pairs
has a relative  frequency of 5.76\% for the same  range of masses (0.8
to  1.4   \msun)  and   separations  (from   1  to   10  kau)   as  in
\citet{Hwang2023}, slightly  larger than  5.35\% quoted in  his paper.
The  number of  astrometric subsystems  with NSS  solutions (AORB  and
ASB1) in  these pairs  is 38,  or 2.5$\pm$0.5\%,  and the  fraction of
astrometric orbits in  the full GCNS for this mass  range is the same,
2.4$\pm$0.1\%.  I conclude that the depletion of astrometric orbits in
wide pairs found by Hwang is  not real, being caused by the inaccurate
Gaia astrometry of stars with astrometric subsystems.

The GCNS hierarchies can clarify  a long-standing issue concerning the
frequency of 2+2 quadruples. Statistical  modeling of the 67 pc sample
of  solar-type stars  \citep{FG67b}  revealed that  presence of  inner
subsystems in both components of  wide pairs is correlated; otherwise,
the number  of predicted 2+2  quadruples would be less  than observed.
Such a conclusion had  been reached  earlier based  on the  presence of
spectroscopic  subsystems in  wide  binaries  \citep{TS2002}.  On  the
other  hand,  a  similar   study  by  \citet{Halbwachs2017}  found  no
correlation between spectroscopic subsystems in components of 116 wide
binaries.

I selected  15,983 pairs of  two stars  wider than 3\arcsec  ~from the
list of 24,606 systems described  above and determined the presence of
subsystems  in  each  component  using either  of  the  Gaia  binarity
indicators  FDBL, RUWE,  and RVERR,  all  with thresholds  of 2.   The
numbers of subsystems  in the primary, secondary,  and both components
are  3542,  2562, and  653,  respectively  (relative frequency  0.222,
0.160, and 0.041).   If the presence of subsystems  in both components
is  uncorrelated, the  expected  frequency of  2+2  quadruples is  the
product  of  subsystem  frequencies   in  primaries  and  secondaries,
$f_{2+2}  =  f_1  f_2  =  0.0355$, while  the  observed  frequency  is
653/15,983=0.0408.  The  excess  of  0.53\%$\pm$0.16\%  is  small  but
formally significant at  the $3.3\sigma$ level.  I  repeated this test
by setting a larger minimum separation  or by using only two criteria,
FDBL  and RUWE,  and obtained  a comparable  excess with  significance
above  $2\sigma$.  In  a  sample  of 116  binaries,  like  the  one  of
\citet{Halbwachs2017},   such   a  small   effect   is   lost  in   the
noise.  Interestingly,  \citet{Fezenko2022}  found  that the simultaneous
presence  of eclipsing  subsystems  in both  components  of Gaia  wide
binaries is significantly enhanced compared  to their frequency in the
field.

\section{Summary and Outlook}
\label{sec:sum}

Combination of  Gaia DR3  data with the ground-based speckle survey
doubles  the  number of  known  hierarchical  systems within  100\,pc,
reaching  now  almost  3,000.   The  estimated total number  of  such
hierarchies is about  20,000, so our current knowledge  is still very
incomplete; some insights on the parameters of missing hierarchies are
given  above.   Several  thousand candidate  nearby  hierarchies  were
extracted from  Gaia, but  for most  of them  the parameters  of inner
subsystems remain  unknown, while the outer  separations are typically
above 100 au.  The main results of this work are:

\begin{enumerate}
\item
A list of 8,032 candidate  hierarchical systems within 100\,pc based on
the GCNS has been created.

\item
A subset of 1,243 candidate hierarchies  brighter than $G = 12$ mag were
observed by  speckle interferometry  at the 4.1  m telescope,  and 506
close inner pairs  were resolved.

\item
New hierarchies  are added to  the MSC,  doubling the number  of known
multiples  within 100\,pc.   The resolved  inner subsystems  and those
with  NSS orbits  occupy different  regions of  the period--mass  ratio
parameter space, with little overlap.

\item
The  amplitudes of  Gaia SB1  and  astrometric orbits  are reduced  by
blending, hence  the mass ratios  derived from those orbits  are lower
limits.

\end{enumerate}

Continued  speckle monitoring  of the  newly discovered  subsystems is
needed for several  reasons.  Orbits of the closest  and fastest inner
pairs with  estimated periods of a  few years can be  determined soon,
filling the  gap between visual and  spectroscopic/astrometric orbits;
the resolved  Gaia orbital  pairs from  Table~\ref{tab:nssspeckle} are
primary  candidates.   Re-observation  of the  remaining  inner  pairs
within  several years  will define  the direction  and speed  of their
orbital  motion.   Its comparison  with  the motion  in the  outer  pairs,
accurately  measured   by  Gaia,  will  give   precious  material  for
the statistical  study  of  relative orbit  orientation  and  eccentricity
distribution; an example of  such analysis for resolved  Gaia triples can
be found  in \citet{Tokovinin2022b}, see also  \citet{Hwang2022a}.  Of
special  interest  will be  a  dynamical  study of  marginally  stable
compact   triples  with   comparable   separations,   like  those   in
Figure~\ref{fig:newtriples}. 

A large sample of  hierarchies with quantified observational selection
is a starting point for inferring the true underlying distributions of
their  parameters   \citep[e.g.][]{FG67b}.   The  limits   of  speckle
detection are  well known, but the SOAR survey covers only  a tiny
fraction of  the GCNS  population.  The  sensitivity of  Gaia binarity
indicators  is still  poorly  understood, although speckle  interferometry
helps in this  respect.  The selection effects in the  NSS catalog are
even more severe.  Despite these obvious difficulties, the prospect of
establishing unbiased  statistics of  binaries and hierarchies  in the
nearby field population is clear.



\begin{acknowledgments}

The research  was funded  by the  NSF's NOIRLab.   This work  used the
SIMBAD   service  operated   by   Centre   des  Donn\'ees   Stellaires
(Strasbourg, France),  bibliographic references from  the Astrophysics
Data System  maintained by  SAO/NASA, and  the Washington  Double Star
Catalog maintained at  USNO.  This work has made use  of data from the
European       Space       Agency       (ESA)       mission       Gaia
(\url{https://www.cosmos.esa.int/gaia}),  processed by  the Gaia  Data
Processing        and         Analysis        Consortium        (DPAC,
\url{https://www.cosmos.esa.int/web/gaia/dpac/consortium}).    Funding
for the DPAC has been provided by national institutions, in particular
the institutions participating in the Gaia Multilateral Agreement.

\end{acknowledgments} 

\facility{SOAR, Gaia}






\end{document}